\documentclass[%
 aip,
 amsmath,amssymb,
 reprint,%
]{revtex4-1}

\usepackage{braket}
\usepackage{graphicx}\usepackage{dcolumn}% Align table columns on decimal point
\usepackage{bm}% bold math
\usepackage{amsmath}
\usepackage[mathlines]{lineno}% Enable numbering of text and display math
\usepackage{esvect}
\usepackage{siunitx}
\usepackage{xcolor}
\usepackage{soul}
\usepackage{makecell}
\usepackage{multirow}

\usepackage[utf8]{inputenc}
\usepackage[T1]{fontenc}
\usepackage{mathptmx}
\usepackage{etoolbox}

\usepackage{hyperref}

\makeatletter
\def\@email#1#2{%
 \endgroup
 \patchcmd{\titleblock@produce}
  {\frontmatter@RRAPformat}
  {\frontmatter@RRAPformat{\produce@RRAP{*#1\href{mailto:#2}{#2}}}\frontmatter@RRAPformat}
  {}{}
}%
\makeatother
\begin{document}

%\preprint{AIP/123-QED}

\title{Collision-induced spectroscopy and radiative association in  microcavities}
\author{Tuan H. Nguyen}
\author{Raphael F. Ribeiro}
\email[]{raphael.ribeiro@emory.edu}
\affiliation{Department of Chemistry and Cherry Emerson Center for Scientific Computation, Emory University, Atlanta, GA, 30322}
\date{\today}

\begin{abstract}
%[Intermolecular collisions are essential in chemical reaction theories, yet their role in reactions under strong coupling with a cavity field has been largely overlooked.] 
Polariton chemistry has emerged as a new approach to directing molecular systems via strong light-matter interactions in confined photonic media. In this work, we implement a classical electrodynamics-molecular dynamics method to investigate collision-induced emission and radiative association in planar microcavities under variable light-matter coupling strength. We focus on the argon-xenon (Ar-Xe) gas mixture as a representative system, simulating collisions coupled to the confined multimode electromagnetic field. We find that while the effects of a microcavity on collision-induced emission spectra are subtle, even at extremely large coupling strengths, radiative association can be significantly enhanced in a microcavity. Our results also indicate that microcavities may be designed to induce changes in the statistical distribution of Ar-Xe complex lifetimes. These findings provide new insights into the control of intermolecular interactions and radiative kinetics with microcavities.
\end{abstract}

\maketitle

\section{Introduction}

\begin{figure}
    \centering
    \includegraphics[width=0.9\linewidth]{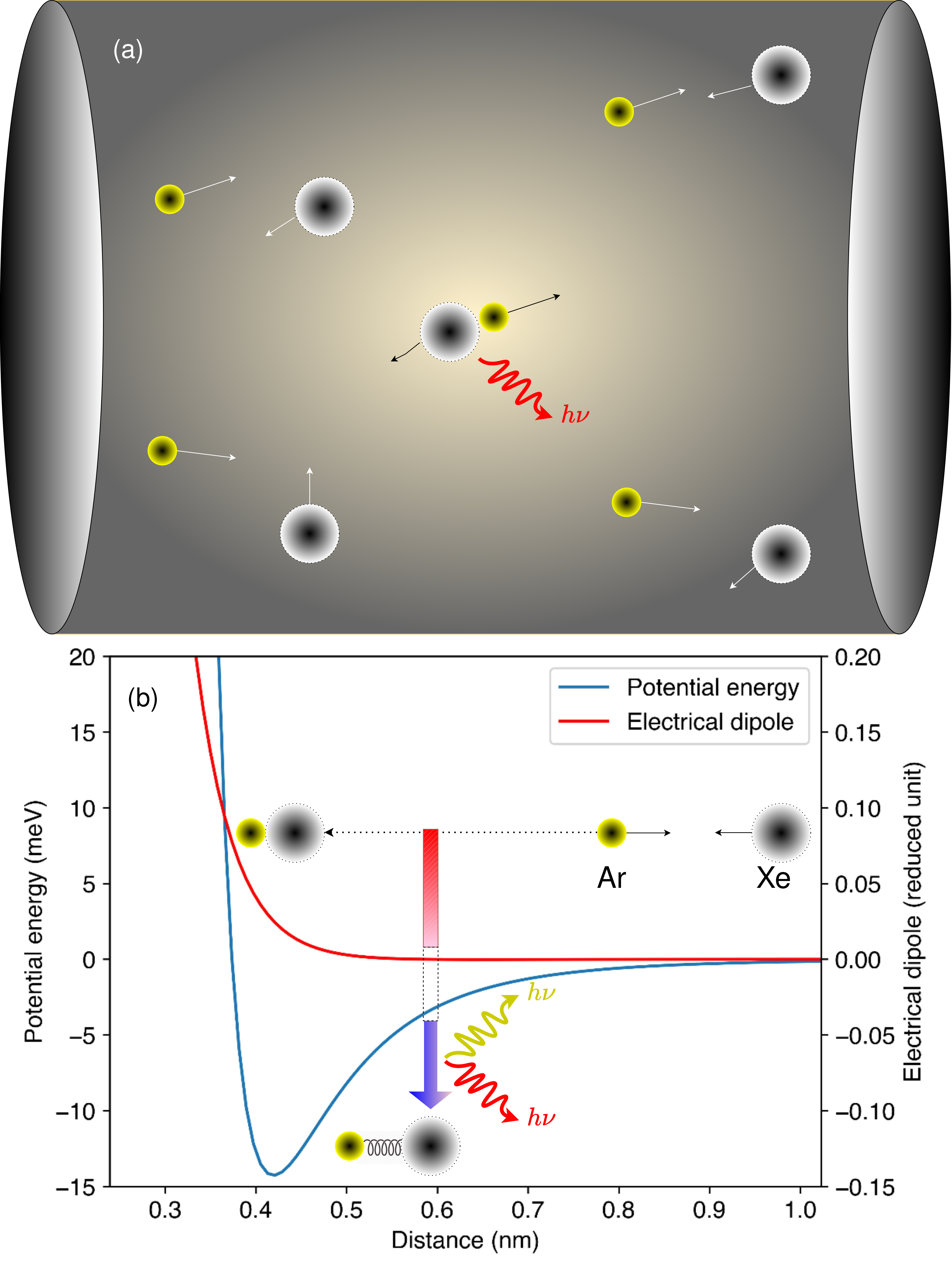}
    \caption{(a) Schematic representation of an Ar–Xe gas mixture confined within a microcavity whose enhanced electromagnetic field couples to transient Ar–Xe dipoles generated during binary collisions. (b) Lennard–Jones potential energy $V(R)$ and dipole function $\mu(R)$ for Ar–Xe, together with an illustration of how photon emission promotes formation of the transient complex.}
    \label{feature_graphics}
\end{figure}

\par Interest in chemistry in microcavities has been fueled by experimental observations of particular chemical reactions suggesting distinct rates and selectivity in optical resonators compared to reactions that occur in free space \cite{thomas2016ground,thomas2019tilting, hirai2020modulation, pang2020role, sau2021modifying, chen2022cavity, ahn2023modification}. A key feature in all studied systems is the strong light-matter interaction regime signaled by the emergence of normal-modes denoted vibrational polaritons which arise from the hybridization of molecular vibrations with microcavity electromagnetic (EM) modes \cite{hopfield1958theory, agranovich1960dispersion, khitrova1999nonlinear, feist2018polaritonic, ribeiro2018polariton, herrera2020molecular, li2022molecular, xiong2023molecular, mandal2023theoretical, weight2023theory, xiang2024molecular, nelson2024more}. Earlier theoretical efforts in polariton chemistry have examined the influence of strong light-matter interactions on various molecular processes including intramolecular vibrational relaxation \cite{chen2022cavity, schafer2022shining}, solute-solvent energy exchange \cite{li2022qm, li2022energy} and chemical equilibria \cite{sun2024theoretical}. However, aside from a few basic studies \cite{cederbaum_making_2024, moiseyev2024complex}, scarce attention has been paid to microcavity effects on atomic collisions, which are fundamental to chemical kinetics \cite{houston2006chemical}. 
%Despite significant advancements in polaritonic chemical theory and modeling, several critical aspects remain underexplored. As discussed earlier, the radiative influence on intermolecular collisions warrants further investigation.

\par Cederbaum et al. \cite{cederbaum_making_2024} recently demonstrated that a strongly confined radiation field can alter $\text{LiH}^+$ formation via scattering of a lithium cation and a hydrogen atom. This work suggests the conversion of collisional energy into EM radiation enhances the probability to form a molecule in a microcavity, though the theory is limited to a pair of atoms and one single cavity mode, whereas strong light-matter interactions are typically accessible with a significant density of molecules \cite{ebbesen2016hybrid, ribeiro2018polariton, wright2023rovibrational} in microcavities with a continuous spectrum corresponding to EM modes with nonvanishing in-plane propagation \cite{kavokin2017microcavities}.

\par The phenomenon examined in Ref. \cite{cederbaum_making_2024} parallels radiative association, a process in dilute astrophysical environments where atomic collisions at low temperatures lead to the formation of molecules via deposition of excess collisional energy into the EM field via emission\cite{nyman2015computational, gustafsson2013classical, szabo_polyatomic_2023, MILFELD1983529}. While typically negligible under weak light-matter coupling in dense environments, radiative mechanisms could play a significant role in optical resonators. In fact, vibrational polariton signatures arise from enhanced radiation-matter interactions \cite{kavokin2017microcavities}. Stronger coupling in infrared resonators could amplify radiative association and dissociation \cite{suyabatmaz2024polaritonic}, potentially unlocking unexplored chemical phenomena.

\par The interaction of light with molecular scattering under different regimes of light-matter interactions is the main focus of this work. Radiative association is inherently linked to collision-induced infrared emission \cite{levine1967classical, meuwly2002dynamical, reguera2006classical}, making it a conceptually suitable context for our exploration. We examine a gaseous mixture of argon (Ar) and xenon (Xe) at various temperatures and non-equilibrium conditions (Fig. \ref{feature_graphics}a). These systems are advantageous for simulations as they have been extensively studied via collision-induced spectroscopy \cite{grigoriev_interaction-induced_1998, gross_collision-induced_2003, gross2003spectroscopic, fakhardji2019collision, fakhardji2021direct}. Importantly, Ar-Xe collisions may result in metastable complexes (Fig. \ref{feature_graphics}b), thereby providing a platform for investigating potential microcavity effects on radiative association. We also note that there has been increasing interest in unraveling microcavity effects on gas-phase chemistry \cite{wright2023rovibrational, wright2023versatile,nelson2024more}.  

\par \textit{Ab initio} quantum electrodynamics\cite{foley2023ab, tancogne2020octopus, schafer2022shining, weight2024cavity, mandal2023theoretical, flick2018strong} and mixed quantum-classical models\cite{luk2017multiscale, zhang2019non, nelson2020non, saller2022accurate} provide natural starting points for examining microcavity effects on reactive molecular systems, but have prohibitive computational cost for simulating time-dependent dynamics involving a large number of molecules and EM modes. Conversely, classical dynamics\cite{miller1978classical} provides a computationally efficient approach that successfully captures polariton formation and dynamics in the collective regime \cite{li_cavity_2020,li2021classicalcavity, li2022qm,li2024mesoscale, schafer2024machine, luk2017multiscale, ji2025selective}. In polaritonic systems where the relevant vibrational modes have sufficiently low frequency (relative to room or any other temperature of interest), classical dynamics is expected to provide a highly accurate description of the polaritonic system, as long as suitable force-fields and dipole functions are employed to model the molecular ensemble.

In this work, we provide a classical electrodynamics-molecular dynamics (MD) investigation of Ar-Xe scattering under different regimes of light-matter interactions in infrared microcavities. Section \ref{sec_theory} describes the employed classical electrodynamics Hamiltonian, equations of motion, force fields, and MD integrator. Section \ref{sec_result} contains Results and Discussion, including a comparison of theoretical and experimental collision-induced emission spectra in free space and simulations of Ar-Xe collision-induced emission and complex lifetime statistics in the microcavity under equilibrium and nonequilibrium conditions. Conclusions are given in Section \ref{sec_conclusion}.

\section{Theory}\label{sec_theory}

\subsection{Classical Light-Matter Dynamics}

In this section, we review the formalism of classical electrodynamics in the context relevant to our work. 

Our model contains an interacting gaseous mixture of Ar and Xe atoms (represented as point-particles) and the dynamic infrared radiation field. We work at low enough temperatures that the atoms and transient complexes can be assumed to be in the adiabatic electronic ground state. When the distance $\mathbf{r}$ between Ar and Xe and the corresponding kinetic energy are sufficiently small, a weakly bound complex (with harmonic frequency much lower than the temperatures of interest to our study) is formed with potential energy curve $V(r)$ and electrical dipole moment $\boldsymbol{\mu}(r)$. Fluctuations in the dipole moment of the Ar-Xe mixture enable it to interact with the radiation field, as probed by collision-induced emission spectroscopy \cite{frommhold1994collision}. Throughout this manuscript, we will only consider systems that are sufficiently dilute that the probability for a three-body collision is negligible and the Lagrangian for the system in the Coulomb gauge can be written in Gaussian-CGS units as
\begin{align}
    L = \sum_\alpha \sum_i \frac{1}{2} m_\alpha \dot{r}_{\alpha,i}^2  - V_{\text{LJ}} + L_{\text{rad}} + \int J_i(\mathbf{r}) A_i(\mathbf{r})\mathrm{d}^3r, \label{eq:lagrangian_c1}
\end{align}
where the subscript $i \in \{x,y,z\}$ denotes the spatial direction, $r_{\alpha,i}$ and $\dot{r}_{\alpha,i}$ are the $i$th component of the position and velocity of atom $\alpha$, respectively; $A_i$ is the $i$th component of the transverse vector potential field $\textbf{A}$, $L_{\text{rad}} \equiv L_{\text{rad}}(\mathbf{A},\dot{\mathbf{A}})$ denotes the Lagrangian of the free radiation field, $V_{\text{LJ}} \equiv V_{\text{LJ}}(\{\mathbf{r}_i\})$ is the Lennard-Jones potential energy of the gas-phase mixture (see below), and $J_i(\mathbf{r})$ is the $i$th component of the matter current density given by
\begin{align}
    J_i = \frac{\partial P_i}{\partial t} \label{eq:jcurrent_1},
\end{align}
where $\mathbf{P}$ is the matter polarization vector field and $P_i$ its $i$th component. 
Note the rotationally invariant light-matter interaction in Eq. \ref{eq:lagrangian_c1}, and the Coulomb gauge condition $\nabla \cdot \mathbf{A} = 0$ implies that only the transverse component of the matter current $\mathbf{J}$ is driven by the radiation field. Let $I_{\text{Xe}}$ and $I_{\text{Ar}}$ be index sets containing all indices for Xe and Ar atoms, respectively. Under the dilute gas-phase condition described above, $P_i$ can be expressed by
\begin{align}
    P_i = \sum_{\alpha \in I_{\text{Xe}}} \sum_{\beta \in I_{\text{Ar}}} \mu_{\alpha\beta,i}\delta(\mathbf{r}-\mathbf{R}_{\alpha\beta}), \label{eq:matter_pol}
\end{align} \normalsize
where $\mu_{\alpha\beta, i} \equiv \mu_i(\mathbf{r}_{\alpha\beta})$ is the $i$th component of the dipole moment vector for the complex consisting of Ar and Xe atoms $\alpha$ and $\beta$ with distance vector $\mathbf{r}_{\alpha\beta}$ and center of mass position $\mathbf{R}_{\alpha\beta}$. Using the polarization density given by Eq. \ref{eq:matter_pol} in the current density expression leads to 
\begin{align}
    J_i = \sum_{\alpha \in I_{\text{Xe}}} \sum_{\beta \in I_{\text{Ar}}} \left[\frac{\partial \mu_{\alpha\beta,i}}{\partial t} + \mu_{\alpha\beta,i} \dot{\mathbf{R}}_{\alpha\beta} \cdot \boldsymbol{\nabla} \right]\delta(\mathbf{r}-\mathbf{R}_{\alpha\beta}),
\end{align}\normalsize
The contribution to the charge density proportional to the  $\alpha\beta$ complex center of mass velocity $\dot{\mathbf{R}}_{\alpha\beta}$ is known as the Rontgen term \cite{craig1998mqed}. It can be safely neglected as the typical speed of the colliding atoms in our simulations is much slower than the speed of light \cite{wilkens1994significance}. In this case, the charge current density can be written as
\begin{align}
   J_i & = \sum_{\alpha \in I_{\text{Xe}}} \sum_{\beta \in I_{\text{Ar}}} \frac{\partial \mu_{\alpha\beta,i}}{\partial t}\delta(\mathbf{r}-\mathbf{R}_{\alpha\beta})
     \nonumber \\
   & \approx  \sum_{\alpha} \frac{\partial \mu_{\alpha,i}}{\partial t}\delta(\mathbf{r}-\mathbf{r}_\alpha) \label{eq:curr_density}
\end{align}
where to obtain the second line we used the assumption that a particular Ar atom is never sufficiently close to more than one Xe so each $\alpha \in I_{\text{Xe}}$ belongs at most to a single transient complex with dipole moment vector denoted by $\boldsymbol{\mu}_{\alpha}$ and that the EM field is constant in the nanoscopic volume occupied by a transient molecule, so we can replace $\mathbf{R}_{\alpha\beta}$ by $\mathbf{r}_\alpha$ (the position of the $\alpha$ atom). 

Inserting the obtained current density expression (Eq. \ref{eq:curr_density}) into the Coulomb gauge Lagrangian (Eq. \ref{eq:lagrangian_c1}) gives
\begin{align}
     L_C = \sum_\alpha \sum_i \frac{1}{2} m_\alpha \dot{r}_{\alpha,i}^2  - V_{\text{LJ}} +  L_{\text{rad}} + \sum_{\alpha}\sum_{i,j}A_i(\mathbf{r}_\alpha)\frac{\partial \mu_{\alpha,i}}{\partial r_{\alpha,j}} \dot{r}_{\alpha,j} \label{eq:lagrangian_c}.
\end{align}

Note that the above Lagrangian strongly resembles the standard Lagrangian for point charges in an EM field \cite{littlejohn_note, craig1998mqed}. The key distinction lies in the spatial gradient of the dipole vector, which serves as effective charges for the dipoles interacting with the EM field. This is further corroborated by the atomic equations of motion
\begin{align}\label{eom_matter}
\begin{split}
    m_\alpha \ddot{r}_{\alpha,j} &= \sum_l \sum_i \Biggr\{  \frac{\partial A_i (\textbf{r}_\alpha)}{\partial r_{\alpha,j}} \frac{\partial \mu_{\alpha,i}}{\partial r_{\alpha,l}} \dot{r}_{\alpha,l} 
    - \frac{\partial A_i (\textbf{r}_\alpha)}{\partial r_{\alpha,l}}  \frac{\partial\mu_{\alpha,i}}{\partial r_{\alpha,j}} \dot{r}_{\alpha,l} 
    \\
    &\quad - \frac{\partial A_i (\textbf{r}_\alpha)}{\partial t} \frac{\partial\mu_{\alpha,i}}{\partial r_{\alpha,j}} \Biggr\} - \frac{\partial V_{\text{LJ}}}{\partial r_{\alpha,j}}.
\end{split}
\end{align}

In both free space and planar microcavity, the EM vector potential $\textbf{A}$ admits the mode decomposition
\begin{align}
    \textbf{A}(\textbf{r}) = \frac{1}{\sqrt{V}}\left[ \sum_{\textbf{k}} \sum_{\lambda} C_{\textbf{k}\lambda} e^{-i\omega t} \textbf{f}_{\textbf{k}\lambda}(\textbf{r}) + C^*_{\textbf{k}\lambda} e^{i\omega t} \textbf{f}^*_{\textbf{k}\lambda}(\textbf{r}) \right], \label{eq:vec_pot}
\end{align}
where $C_{\textbf{k}\lambda}$ is the time-dependent amplitude of the mode with wavevector $\mathbf{k}$ and polarization $\lambda \in \{1,2\}$, $V$ is the volume, $\omega = c |\textbf{k}|$ is the frequency with $c$ being the speed of light and  $\textbf{f}_{\textbf{k}\lambda}$ is the mode profile vector determined by boundary conditions. In free space modeled as a 3-torus (cube with periodic boundary conditions) with length $L$, and volume $V = L^3$, the field modes are described by $\mathbf{f}_{\textbf{k}\lambda} (\textbf{r}) =  e^{i\textbf{k} \cdot \textbf{r}}\boldsymbol{\epsilon}_{\textbf{k}\lambda}$ with $\boldsymbol{\epsilon}_{\textbf{k}\lambda}, \lambda = 1,2$ corresponding to linearly independent polarization vectors transverse to $\textbf{k}$. In a planar microcavity with confinement along the $z$ axis by perfect metallic mirrors, the EM field modes are given by \cite{hinds_cavity_1990}:
\begin{align}
\begin{split}
    \textbf{f}^{\text{TE}}_{\mathbf{k}}(\textbf{r}) &= \sin (k_z z)  e^{i \mathbf{k}_{\parallel} \cdot\mathbf{r} } \mathbf{n}_{\mathbf{k}_{\parallel}} \times \mathbf{n}_z, \\
    \textbf{f}^{\text{TM}}_{\mathbf{k}}(\textbf{r}) &= \frac{|\textbf{k}_{\parallel}|}{|\textbf{k}|} \cos(k_z z) e^{i \mathbf{k}_{\parallel} \cdot \textbf{r}} \textbf{n}_z - i \frac{k_z}{|\textbf{k}|} \sin(k_z z) e^{i \textbf{k}_{\parallel} \cdot \textbf{r}} \textbf{n}_{\mathbf{k}_{\parallel}},
\end{split}
\end{align}
where TE denotes transverse-electric polarization and TM denotes transverse-magnetic polarization, $\mathbf{k}_{\parallel} = (k_x, k_y, 0)$ is the in-plane wave vector, $\textbf{n}_{\mathbf{k}_{\parallel}}, \textbf{n}_z$ are unit vectors along the $k_\parallel$ and $z$ directions, respectively, and $|\mathbf{x}|$ denotes the Euclidean norm of vector $\mathbf{x}$. 

Thus, propagating $\textbf{A}$ is tantamount to solving for the time-dependent field amplitudes ${C_{\textbf{k}\lambda}}$. Their equations of motion are given by
\begin{align}\label{eom_field_eq}
    \dot{C}_{\textbf{k}\lambda} &= \frac{2\pi i c}{\omega \sqrt{V}} e^{i\omega t} \int \textbf{J}(\textbf{r}) \cdot \textbf{f}^*_{k\lambda}(\textbf{r}) d^3 \textbf{r}
    \end{align}
    
This is a linear equation that can be readily solved given the current density $\mathbf{J}$ 
\begin{align}\label{eom_field}
    C_{\textbf{k}\lambda}(t) &= C_{\mathbf{k}\lambda}(0)+ \frac{2\pi i c}{\omega \sqrt{V}} \sum_\alpha \sum_{i,j} \int^{t}_{0}  \frac{\partial \mu_{\alpha,i}}{\partial r_{\alpha,j}} \dot{r}_{\alpha,j} f^*_{k\lambda,i} (\textbf{r}_\alpha) e^{i\omega t'} \; dt',
\end{align}
where $f_{k\lambda, i}$ is the projection of the mode function along the $i$th direction. 

 Eq.\ref{eom_field} can be substituted into Eq.\ref{eq:vec_pot}, and subsequently this expression inserted into Eq.~\ref{eom_matter}, to derive an equation of motion for the atomic system solely in terms of molecular quantities and the initial conditions of the electromagnetic field. In other words, the EM field is integrated out exactly, resulting in a dynamical equation for the atomic system wherein EM degrees of freedom no longer explicitly appear. This formulation yields an integro-differential equation for the atomic dynamics, with explicit dependence on the system's prior history at all earlier times $0 \leq t' < t$, but it also discards the electromagnetic degrees of freedom, preventing a direct probe of the field microstate evolution as required for a description e.g., of field thermalization (see Sec. III.D). Therefore, we keep the mode amplitudes and integrate the equivalent set of first-order, local-in-time equations with the time-domain solver detailed in the next section.

\subsection{Molecular Dynamics Integrator}

We obtain the dynamical evolution of the mixed light-matter system by numerically integrating Eqs. \ref{eom_matter} and \ref{eom_field}. We employ the following scheme:

\begin{align}
&    \textbf{v}_{\alpha,t+\Delta t/2} = \textbf{v}_{\alpha,t} + \frac{\Delta t}{2m_\alpha} \mathbf{F}_\alpha(t, \textbf{r}_{\alpha,t}, \textbf{v}_{\alpha,t}, \textbf{C}_t),
    \nonumber\\
 &   \textbf{r}_{\alpha,t+\Delta t} = \textbf{r}_{\alpha,t} + \textbf{v}_{\alpha,t+\Delta t/2} \Delta t,
    \nonumber\\
  &  \textbf{v}_{\alpha,t+\Delta t}= \textbf{v}_{\alpha,t+\Delta t/2} + \frac{\Delta t}{2m_\alpha} \mathbf{F}_\alpha(t+\Delta t, \textbf{r}_{\alpha,t+\Delta t}, \textbf{v}_{\alpha,t + \Delta t/2}, \textbf{C}_t),
    \nonumber \\
  &  \textbf{C}_{t+\Delta t} = \textbf{C}_t + \textbf{G}(t+\Delta t,\textbf{r}_{\alpha,t+\Delta t}, \textbf{v}_{\alpha,t + \Delta t}) \Delta t,
\end{align}
where $\textbf{F}_\alpha$ is the force exerted on atom $\alpha$ as given in Eq. \ref{eom_matter}, and $\textbf{G}$ abbreviates the right-hand-side of the field amplitude equation of motion (Eq. \ref{eom_field}). This scheme combines two symplectic integration methods: the velocity Verlet algorithm\cite{swope1982computer} is applied to propagate the positions and velocities of the atoms in the presence of the field, and the symplectic Euler \cite{cromer1981stable} algorithm is used to evolve the field amplitudes. A demonstration of the quasi-energy conservation of this algorithm is given in the Supporting Information.

We apply periodic boundary conditions to model the free space radiation field, whereas the microcavity is confined along the $z$ axis and periodic along the $x$ and $y$ directions. This leads to quantized electromagnetic mode wavevectors $\mathbf{k} = (k_x,k_y,k_z)$ satisfying
\begin{align}
\begin{split}
    \mathbf{k} &= \begin{cases}
        \left(2\pi n_x/L_x, 2\pi n_y/L_y, 2\pi n_z/L_z \right), \;\, (\text{free-space})
        \\
        \left(2\pi n_x/L_x, 2\pi n_y/L_y, \pi n_z/L_z \right), \quad (\text{microcavity})
    \end{cases}
\end{split}
\end{align}
where $L_x, L_y, L_z$ are the lengths of the box along the x,y,z directions. In all simulations, we selected $L_x = L_y$, while $L_z$ was adjusted based on the modeled scenario (e.g., inside or outside a microcavity). The atomic subsystem is treated with periodic boundary conditions in free space and within a microcavity.

\subsection{Simulation Details}
Ar-Xe collisions have been experimentally shown to produce radiation at exceptionally low wavenumbers (approximately $50  ~\text{cm}^{-1}$) \cite{grigoriev_interaction-induced_1998}. To reliably simulate such low-frequency modes, the required box dimension extends up to 1 cm (note $\Delta k_i \propto 1/L_i$ and $\omega = c|\mathbf{k}|$). To mitigate computational challenges associated with simulating such a large system, we adopted a Monte Carlo-Molecular Dynamics approach discussed next.

The initial positions of the Ar atoms are uniformly distributed within the simulation box. Each Ar atom defines the center of a sphere with a radius of 5 reduced units (approximately $17.05 \si{\angstrom}$). A single Xe atom is placed on the surface of each such sphere. The velocities of both Ar and Xe atoms are sampled from a Maxwell-Boltzmann distribution, with their directions randomly sampled within a certain angle toward each other to ensure a collision (with a non-vanishing transient Ar-Xe dipole moment; see additional details in Supporting Information). The simulation ends when all Ar and Xe atoms are sufficiently far from each other that the total potential energy drops below $10^{-4}$ reduced units (see Supporting Information) or the number of timesteps exceeds $10^4$.

This setup allows us to explore light-matter interactions in a dilute Ar-Xe gas mixture, where three-body collisions are negligible. Ar-Ar and Xe-Xe collisions, lacking significant transient dipoles, are excluded; we focus exclusively on relevant Ar-Xe collisions. To efficiently capture collision statistics, our simulations start with Ar-Xe pairs initially separated by approximately 2 nm, far beyond the Ar-Xe van der Waals potential range ($\approx  0.3 ~\text{nm})$. Each pair is assigned random initial velocities and impact parameters, ensuring sequential rather than simultaneous collisions. Thus, although initialized with multiple Ar-Xe pairs, our simulations accurately model isolated pair collisions within a dilute gas-phase environment. This approach is validated numerically through comparisons with experimental data in free space conditions (see Sec. III.A).

We employ the Lennard-Jones potential with parameters for the Ar-Xe gas mixture from a prior theoretical study \cite{gross_collision-induced_2003}: 
\begin{align}\label{lennardjones}
    V_{LJ}(r_{\alpha\beta}) = 4\varepsilon_{\alpha\beta} \left(\frac{\sigma_{\alpha\beta}^{12}}{r_{\alpha\beta}^{12}} - \frac{\sigma_{\alpha\beta}^6}{r_{\alpha\beta}^{6}}\right)
\end{align}
with $r_{\alpha\beta}$ is the distance between $\alpha$ and $\beta$, and $\sigma_{\alpha\beta}, \varepsilon_{\alpha\beta}$ are Lennard-Jones parameters for Argon-Argon, Xenon-Xenon, or Argon-Xenon interactions. The potential energy for the Ar-Xe complex is provided in Fig. \ref{feature_graphics}b.

Throughout, we employ the reduced units system where length, energy, and mass units are defined by $\sigma_\text{{Ar-Ar}}, \varepsilon_{\text{Ar-Ar}}$, and $m_{\text{Ar}}$, respectively. The reduced units are abbreviated as r.u. in the remaining text. The values for all parameters in reduced and CGS units are included in the Supporting Information.

To model the light-matter interaction, we utilize the Ar-Xe dipole function derived from published experimental data \cite{grigoriev_interaction-induced_1998}. In particular, this dipole function was measured for a relatively dilute gas, minimizing three-body interactions compared to other studies and making it particularly well suited for our simulations which operate under extremely dilute conditions, e.g., they contain approximately 1,000 atoms within a $\sim 1 ~\text{cm}^3$ box. Consequently, the probability of two Ar-Xe pairs being close enough for a three or four-body interaction is astronomically small. The functional form of the Ar-Xe dipole vector $\boldsymbol{\mu}$ is provided below (see also Fig. \ref{feature_graphics}b).
\begin{align}
    \boldsymbol{\mu} (\textbf{r}_{\text{Xe}}- \textbf{r}_{\text{Ar}}) &= \left[\mu_0 e^{-a (R - R_0)} - \frac{D_7}{R^7} \right] \frac{\mathbf{r}_{\text{Xe}} -\mathbf{r}_{\text{Ar}}}{R},
\end{align}
where $\mu_0, a, R_0$, and $D_7$ are real positive scalar parameters and $R =|\textbf{r}_{\text{Xe}} - \textbf{r}_{\text{Ar}}|$ is the Euclidean distance between a particular Ar and Xe atom.

In general, calculating the dipole, the dipole gradient, the interatomic potential, and the forces at each time step requires exhaustive evaluation of all pairwise atomic distances and vectors. To streamline this process, we employ the cell list technique \cite{frenkel2023understanding}, which efficiently skips distance calculations for atom pairs separated by distances exceeding a predefined threshold.

The initial conditions for the EM field modes are sampled from the classical Boltzmann distribution describing the radiation field represented by the complex amplitudes $C_{\mathbf{k}\lambda} = |C_{\textbf{k}\lambda}| \exp(i\phi_{\mathbf{k}\lambda})$. In terms of the magnitude and phase dynamical variables $|C_{\textbf{k},\mu}|$ and $\phi_{\mathbf{k}\mu}$ , the thermal distribution is given by
\begin{align} \label{amplitude_sampling}
   & f(\phi_{\mathbf{k}\lambda})d\phi_{\mathbf{k}\lambda} = \frac{1}{2\pi} d\phi_{\mathbf{k}\lambda}, \\
    &f(|C_{\mathbf{k}\lambda}| )d|C_{\mathbf{k}\lambda}| = \frac{\omega_\mathbf{k}^2 |C_{\mathbf{k}\lambda}|}{\pi c^2 k_BT} \exp\left[-\frac{\omega_\textbf{k}^2}{\pi c^2 k_BT} \frac{|C_{\mathbf{k}\lambda}|^2 }{2} \right] d|C_{\mathbf{k}\lambda}|.
\end{align}
In words, the angles $\phi_{\mathbf{k}\lambda}$ are sampled from a uniform distribution, while the magnitudes $|C_{\textbf{k}\lambda}|$ are sampled from the $\chi_k$-distribution with $k = 2$.

\par We directly compute collision-induced infrared spectra of the Ar-Xe mixture from the time-dependent fluctuations in the EM field energy profile. The energy of the field at time $t$ assuming a discrete mode spectrum can be written as $E_\text{EM}(t) =  \sum_{\omega \geq 0} E_\omega(t)$, where $E_\omega(t)$ is the energy stored in all modes with frequency $\omega$ as given by
\begin{align}
    E_\omega(t) = \frac{\omega^2}{2\pi c^2} \sum_{\textbf{k} \in S^2(\omega / c)} \sum_{\lambda=1}^{2} | C_{\textbf{k}\lambda}(t) |^2,
\end{align}
where $S^2(\omega/c)$ is the set containing all wave vectors $\mathbf{k}$ with length $|\mathbf{k}| = \omega/c$. The time-averaged fluctuations of $E_{\omega}(t)$ are proportional to the collision-induced EM spectrum. Specifically, let $E_\omega(t_1), E_\omega(t_2), \ldots, E_\omega(t_N)$ be the energy at $N$ time points during the simulation and $\bar{E}_\omega$ be the time-average. The lineshape for collision-induced radiation can be expressed by
\begin{align}
    I(\omega) \equiv \sqrt{ \sum_{n=1}^N \frac{1}{N} \bigg[ E_{\omega}(t_n) - \bar{E}_\omega \bigg]^2},
\end{align}
where $\overline{E}_\omega$ is the time average of $E_\omega$ (See Supplementary Material for derivation).

\par 
All simulations in this study probe long-wavelength physics at temperatures below $300\,\mathrm{K}$.  
The radiation field is band-limited between $0.5\,\mathrm{cm}^{-1}$ and $300\,\mathrm{cm}^{-1}$.  
Consequently, mass renormalization effects \cite{spohn2004dynamics, ford1991radiation} arising from our nonperturbative light--matter treatment are negligible and have no impact on the dynamics, whether in free space or within a microcavity.

%\par Figure \ref{freespec256} shows that our simulations following the protocols prescribed in this section reliably capture the collision-induced spectrum of Ar-Xe in free space.

\section{Results and Discussion}\label{sec_result}
\subsection{Free-space Collision-Induced Spectra}\label{subsec_CIE}

In this section, we benchmark our computational approach against experimental collision‐induced spectra obtained for Ar–Xe under low-pressure conditions in free space. In this case, the radiation-matter interaction is extremely weak, and emission into the radiation continuum is effectively irreversible so the EM field exerts no appreciable back-action on the gas. Consequently, the corresponding collision-induced spectra can be accurately reproduced by the power spectrum of the dipole current correlation function (Larmor's formula \cite{larmor1897lxiii}) without explicitly propagating EM degrees of freedom \cite{grigoriev_interaction-induced_1998, reguera2006classical,gustafsson2013classical,fakhardji2019collision,fakhardji2021direct}. Reflecting this physical picture, in this section, we initialize every EM mode at $T = 0\,\mathrm{K}$, as this choice suppresses simulation noise yet preserves the essential dynamics of collision-induced emission. Simulations that start with the radiation field in thermal equilibrium with the gas at 100 and 292\,K produce broader and greater intensity lines in accordance with the corresponding classical radiation field thermal distribution, yet all salient spectral features remain intact (see detailed discussion at Supplementary Information, Sec.~5).

The noble-gas mixture itself is held at variable temperatures, including that (292\, K) experimentally studied by Grigoriev et al \cite{grigoriev_interaction-induced_1998}. The system is located in a cubic box of side
$L_x = L_y = L_z = 3 \times 10^{7}\,\text{r.u.} \approx 1\,\text{cm}$. The EM field comprises all
modes with wave-number magnitude $k$ from $1$ to $253\,\text{cm}^{-1}$, sampled at $\approx 1\,\text{cm}^{-1}$ intervals. For simplicity, only wave vectors with $k_x \neq 0$ are included, for with a sufficiently large ensemble of randomly oriented collisions, the averaged collision-induced spectrum is effectively isotropic.

\begin{figure}
    \centering
    \includegraphics[width=1.0\linewidth]{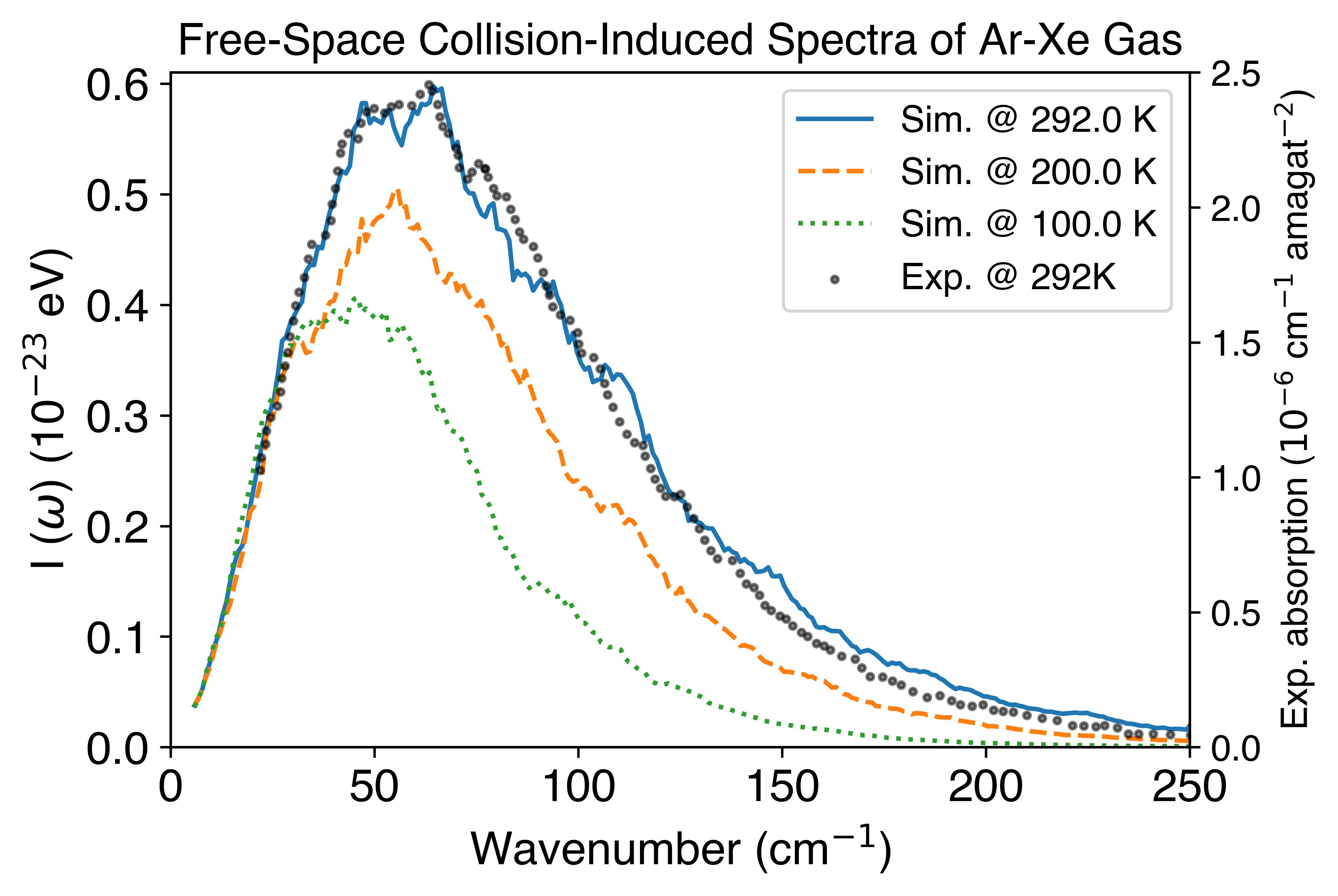}
    \caption{Theoretical (Sim.) collision-induced emission spectra of Ar-Xe gas mixture from classical molecular electrodynamics simulations following the prescription in Sec. II, and experimental (Exp.) absorption spectrum in free-space from Ref. \cite{grigoriev_interaction-induced_1998}. The visible agreement between theory and experiment validates our methodology.}
    \label{freespec256}
\end{figure}

We prepare the system with 512 randomly generated Ar-Xe collision pairs in the presence of the free space EM field. We conducted 20 simulations with randomly sampled atomic positions and velocities, as discussed in Section II. Figure \ref{freespec256} shows the obtained Ar-Xe collisional spectra at various temperatures, averaged over all replicas, alongside experimental spectra at 292 K from Ref \cite{grigoriev_interaction-induced_1998}. 

The results show that the simulated lineshape is quantitatively accurate, validating both our theory and computational methodology. The broad lineshape of the collision-induced spectrum is often attributed to the short finite lifetime of the collision-induced dipole \cite{fakhardji2021direct}. The Ar-Xe collision-induced spectral lineshape is also likely to be influenced by factors such as the distribution of collisional speed, impact parameters, and spatial configuration. Consequently, one might expect the lineshape to also display inhomogeneous broadening. However, as we show next, inhomogeneous broadening due to the aforementioned factors plays a negligible role in the spectra given in Fig. \ref{freespec256}.

\begin{figure}
    \centering
    \includegraphics[width=1.0\linewidth]{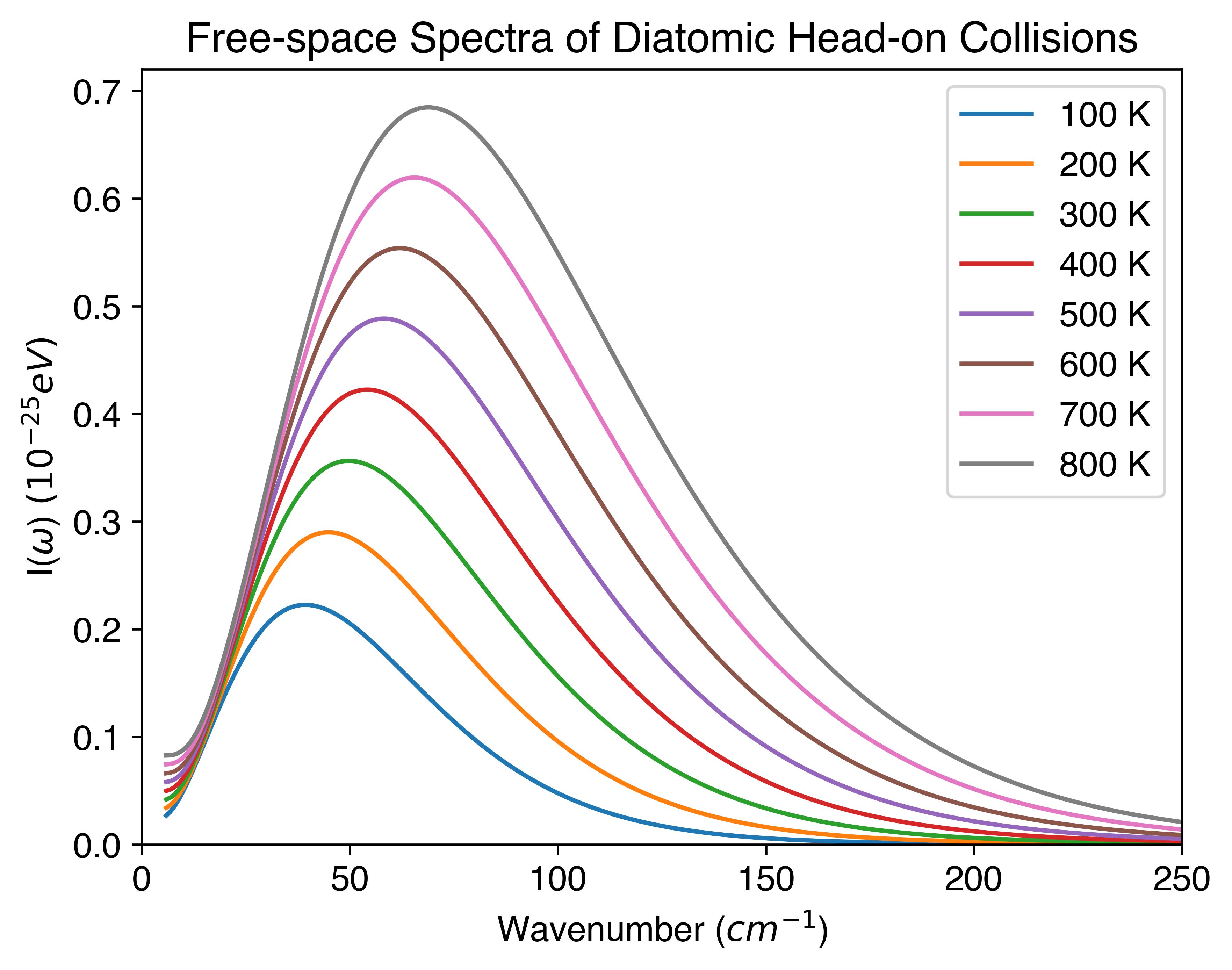}
    \caption{Free-space infrared spectra obtained for collinear (head-on) diatomic Ar-Xe collisions with various total energies in K. The broad lineshapes obtained even at low kinetic energies indicate that homogeneous broadening is the essential mechanism for the line broadening observed in collision-induced spectra of the Ar-Xe mixture.}
    \label{freespec_single}
\end{figure}

We further examined the mechanism underlying the broad collision-induced lineshape observed in Fig. \ref{freespec256} by simulating head-on (collinear collisions) between a single Ar and Xe atoms with equal kinetic energy. Using the same initial conditions and field modes (with nonvanishing $k_x$) as in prior simulations, we orient the collisions along the $z$-axis to maximize the light-matter coupling. Figure \ref{freespec_single} shows the obtained collision-induced spectra for various total kinetic energies (measured in K). The results imply the collision-induced emission lineshape is intrinsically as broad as that obtained for a gaseous mixture (Fig. \ref{freespec256}), leaving no room for inhomogeneous broadening due to variable initial conditions. Overall, the results in this section strongly suggest that a multimode description of the EM field is essential for studying the effect of enhanced light-matter coupling strength on molecular scattering.

\subsection{Microcavity Collision-Induced Spectra}

%Note that inside a microcavity, transverse magnetic and transverse electric modes with nonvanishing $k_z$ have a minimum value   $\pi/L_z = 488.8 \text{ cm}^{-1}$  much larger than the wavenumber range relevant to collision-induced emission (Fig. \ref{freespec256}). Therefore, only TM0 modes $\mathbf{k} = (k_x,k_y,0)$ are coupled with the collision and transient Ar-Xe molecules. 

In this section, we describe our simulations of collision-induced spectra in a microcavity. To emulate the microcavity environment, we reduce the length along the z-dimension to $L_z = 3 \times 10^3 ~\text{r.u.} \approx 1~\mu m$, while keeping the $x$ and $y$ dimensions fixed at the same values as in free-space. Note that inside a microcavity, transverse magnetic and transverse electric modes with nonvanishing $k_z$ have a minimum value   $\pi/L_z = 488.8 \text{ cm}^{-1}$  much larger than the wavenumber range relevant to collision-induced emission (Fig. \ref{freespec256}). 
This indicates that coupling of the collision-induced dipole to the TM$_1$ and TE$_1$ cavity modes \cite{barnes2020classical} is negligible compared to coupling with TM0 \cite{barton1970quantum} modes having frequencies within the $1-253~\text{cm}^{-1}$ interval. Consequently, we omit the TE$_1$ and TM$_1$ modes from our considerations and focus exclusively on TM0 modes with $\mathbf{k} = (k_x, k_y, 0)$ in the frequency range relevant to the Ar-Xe dipole response.
Since there is no distinction between the $x$ and $y$ directions, we include only the modes in TM0 with $k_y = 0$ and $k_x \neq 0$, consistent with our analysis in free space. This cavity geometry will be used for the rest of this work. 

%The rescaled dipole functions allow us to consider the effects of enhanced light-matter coupling strength on the Ar-Xe mixture that may be achievable in other photonic systems where the light-matter coupling enhancement relative to free space may be substantially larger than is the case with a Fabry-Perot microcavity.
%We simulate 512 pairs of colliding Ar and Xe atoms, applying the dipole scaling $\mu(r) \rightarrow \gamma \mu(r)$ with $\gamma = 10 \overline{L}$ and $\gamma = 20\overline{L}$ \textcolor{blue}{($\overline{L} = L_x/\sigma$ is the value of dimension of the simulated box along the x direction in reduced unit)} to compensate for the macroscopic volume of the simulated box and the dilute conditions imposed by computational limitations. 

We examined Ar-Xe collisions at 100 K strongly coupled to the microcavity radiation field at the same temperature.
We simulate 512 pairs of colliding Ar and Xe atoms, applying the dipole scaling $\mu(r) \rightarrow \gamma \mu(r)$ with $\gamma = 10\overline{L}$ and $\gamma = 20\overline{L}$. Here, $\overline{L} = L_x/\sigma$ is the reduced dimension of the simulation box along the $x$-direction.

\par The Ar-Xe dipole magnitude was rescaled, for the free space and microcavity collision induced spectra are indistinguishable when $\gamma = 1$ due to the large mode volume of planar microcavities such as those examined here.
The scaling factor $\gamma$ offsets the large $x$ and $y$ dimensions of the system, effectively allowing us to mimic a significantly smaller mode volume in comparison to typical Fabry-Perot microcavities employed in vibrational strong coupling experiments. Alternatively, the scaled dipole could be viewed to arise from a collective interaction with a large density of simultaneous collisions, however, consideration of effects associated with increased density, such as three-body collisions and pressure broadening, is outside the scope of the current study.  Nonetheless, as mentioned, rescaling the dipole allows us to explore the effects of enhanced light-matter coupling strength on the Ar-Xe mixture that may be attainable in other photonic systems where radiative effects may be amplified due to a reduction in the mode volume.
Later in section \ref{subsec_polariton}, we show that at very low temperatures and special initial conditions, the Ar-Xe complex exhibits unambiguous Rabi oscillations and a polaritonic spectrum when the dipole scaling factors are chosen as described above.

\par
\begin{figure}
    \centering
    \includegraphics[width=1.0\linewidth]{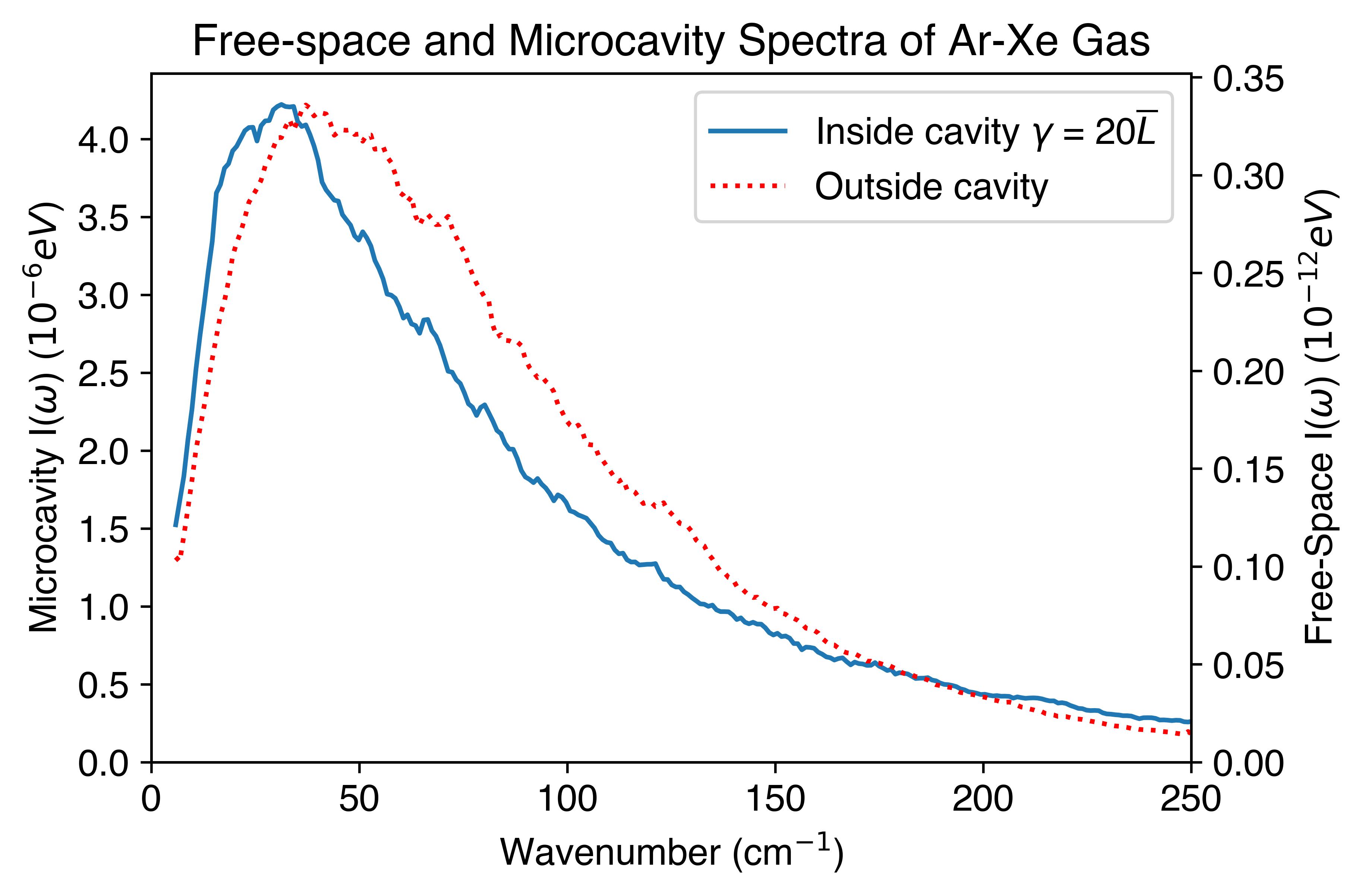}
    \caption{Collision-induced spectrum of Ar - Xe gas mixture at 100K in free space and in a microcavity at the same temperature. The microcavity simulations are distinct from free space due to the field confinement along the $z$ axis (see text) and by the introduction of a dipole scaling factor $\gamma = 20\overline{L}$, which further enhances the light-matter coupling strength. The main difference between the free space and microcavity spectra is in the low-frequency range, as the maximum in the microcavity collision-induced spectrum is redshifted relative to free space and approaches $23~\text{cm}^{-1}$. This frequency corresponds to the natural harmonic frequency of transient Ar-Xe molecules generated via radiative association.}
    \label{cavespec_b}
\end{figure}

 We conducted 40 simulation replicas, totaling 20,480 Ar-Xe collisions. The averaged collision-induced spectrum for the microcavity with $\gamma = 20\overline{L}$ is shown in Fig. \ref{cavespec_b}, while the corresponding spectrum for $\gamma = 10\overline{L}$, which is very similar, is provided in the Supporting Information for reference. We also demonstrated that our calculation has converged with respect to $L_x$  in Fig. S6 and S7 of the SI. A notable distinction between the collision-induced spectra outside and inside a microcavity with enhanced light-matter coupling (Fig. \ref{cavespec_b}) is the dramatic increase in the scale of EM energy fluctuations by orders of magnitude. This follows from Eq \ref{eom_field} which shows that the dipole-driven component of the field amplitude $C_{\textbf{k}\lambda}$ is amplified by a factor $\gamma \sqrt{V_{\text{free}} / V_{\text{cavity}}} \approx 10^{10}$ in the microcavity setup.

\par While an increase in the radiation field temperature inherently amplifies fluctuations in the collision-induced spectrum, the primary cause for the enhanced magnitude of fluctuations observed in Fig.~\ref{cavespec_b} is the strengthened light-matter interactions within the microcavity. This conclusion is further supported by SI Fig.~S11, which demonstrates that similarly large fluctuations are already present in microcavity collision-induced spectra even when the radiation field is initialized at 0K, matching the conditions used in Fig.~1.

%As we show below, the stronger low-frequency intensity fluctuations relative to free space are due to enhanced Ar-Xe collisions leading to association, i.e., transient Ar-Xe complex formation.
The broad linewidths of the simulated collision-induced spectra prevent the emergence of any polaritonic feature in Fig. \ref{cavespec_b}. Nevertheless, the enhanced light-matter coupling strength markedly changes the dynamics of Ar-Xe collisions, as evidenced by the distinct collision-induced spectra obtained for gas mixtures in and out of a microcavity (Fig. \ref{cavespec_b}). 
As we show below, the stronger low-frequency intensity fluctuations relative to free space are due to associative Ar-Xe collisions, i.e., transient Ar-Xe complex formation. 
This occurs when low kinetic energy Ar-Xe collisions with low impact parameters lead to the formation of a transient weakly bound diatomic molecule. Near the minimum of the Ar-Xe Lennard-Jones potential (Eq. \ref{lennardjones}), this system vibrates with frequency $\approx 23 ~\text{cm}^{-1}$, leading to a significant new feature in the collision-induced emission spectrum around this frequency. We pursue a detailed description of the effect of large light-matter coupling strength on the Ar-Xe complex formation dynamics with the trajectory analysis below. 
 
\subsection{Microcavity Effects on Ar-Xe Lifetime Distribution}
\begin{figure}
    \centering
    \includegraphics[width=1.0\linewidth]{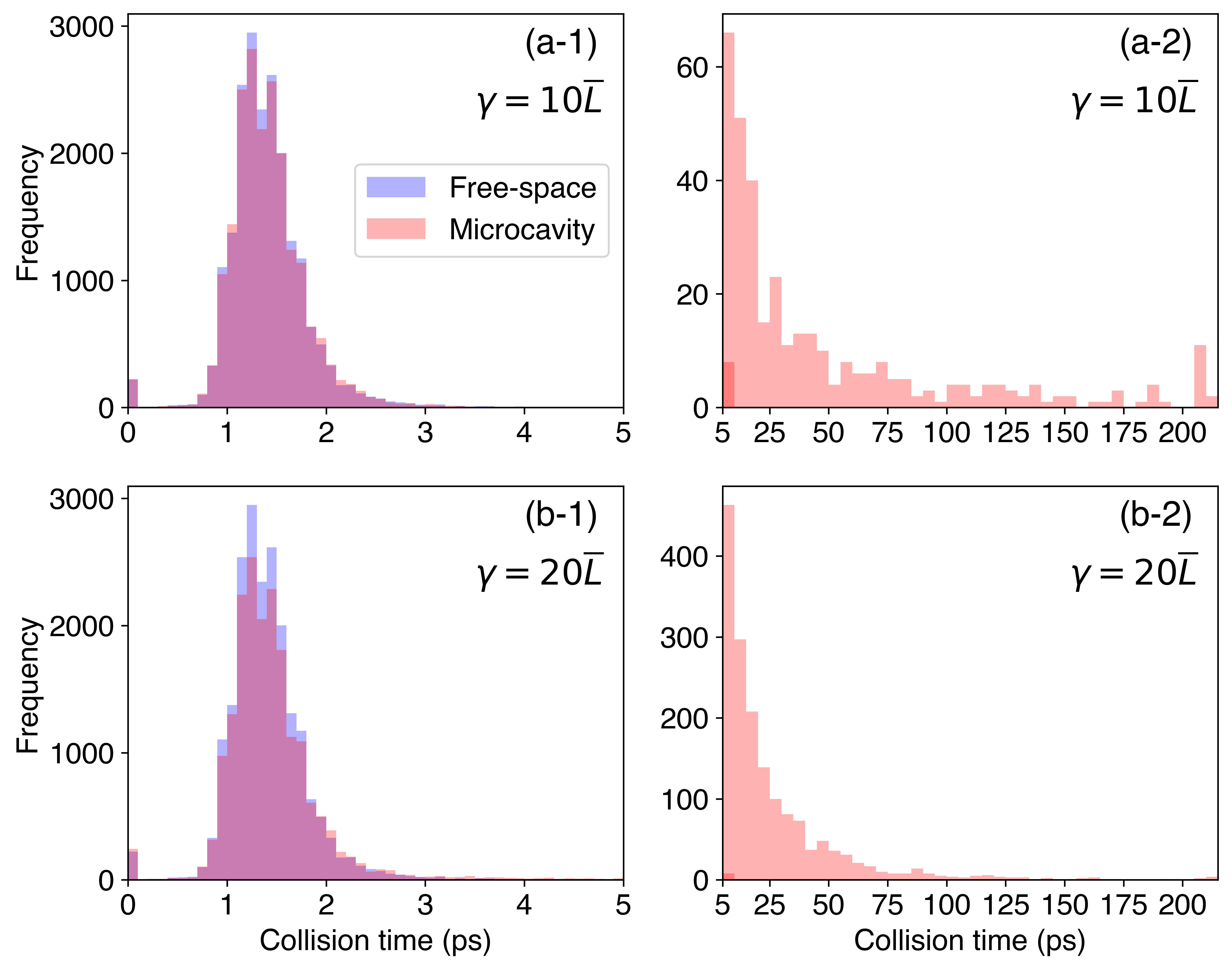}
    \caption{Histograms of Ar-Xe collision times in microcavity environments (with dipole scaling factors $\gamma = 10\overline{L}$ in the first row and $\gamma = 20\overline{L}$ in the second row) compared with free space conditions (left column only). The left column (1) shows histograms with 0.1 ps bin widths covering the full range of collision times, while the right column (2) uses 3 ps bin widths to highlight long-duration collisions (the right tails of the distributions in a-1 and b-1). In our simulations in a microcavity, multiple Ar-Xe pairs showed collision times far exceeding 10 ps, indicating molecular association. By contrast, in free-space simulations, no such association was observed, and all collisions lasted for less than 10 ps.}
    \label{histogram_b}
\end{figure}

\begin{table*}[]
    \centering
        \begin{tabular}{|c|c|c|c|c|c|c|c|}
        \hline 
        \\[-1em]
        & $\text{T}^\text{field}_i$(K) & $\text{T}^\text{field}_f$(K) & $\text{N}_{\text{Ar-Xe pairs}}$ & $\gamma$ ($\overline{L}$) & $\overline{\tau}$ (ps) &  $ \eta_{\tau\geq 10}$ (\%) & $\eta'_{\tau\geq 200}$  (\textperthousand)
        \\ \hline
        \hline
        \multirow{2}{5em}{Equilibrium} & 100  & $104 \pm 5$ & 512 & 10 & 57.77 & 1.37 & 0.63
        \\ 
        & 100 & $106 \pm 4$ & 512 & 20 & 33.01 & 5.79 & 0.29
        \\ \hline
        \hline
        \multirow{2}{5em}{\makecell{Non - \\ equilibrium}} & 0 & $14 \pm 2$ & 256 & 20 & 58.36 & 4.96 & 3.08
        \\
        & 0 & $18 \pm 2$ & 512 & 20 & 45.60 & 5.88 & 1.71
        \\ \hline
        \end{tabular}
  \caption{Statistical data for Ar-Xe collision simulations in a microcavity. Parameters include initial and final EM field temperatures ($T^\text{field}_i$ and $T^\text{field}_f$), number of Ar-Xe pairs per replica ($\text{N}_{\text{Ar-Xe pairs}}$), average lifetime of Ar-Xe complexes ($\overline{\tau}$), percentage of collisions lasting longer than 10 ps ($\eta_{\tau \geq 10}$), and per mille of collisions lasting $\geq 200$ ps ($\eta'_{\tau \geq 200}$). Data represent averages across multiple simulation replicas, totaling up to 20,480 Ar-Xe collisions per case. Results indicate that higher coupling strengths or increased numbers of Ar-Xe pairs enhance the formation rate of transient complexes while reducing their lifetime. Additionally, Ar-Xe complexes formed in a 0K microcavity demonstrate significantly greater stability compared to those in a 100K microcavity.}
    \label{table_summary}
\end{table*}

To assess the impact of enhanced radiative coupling strength in the dynamics of Ar-Xe complexes, we examined the statistics of collision times. The Ar-Xe collision time is defined as the duration in which the corresponding dipole magnitude $|\boldsymbol{\mu}(R)| > 0 $. In free space, even at $T = 100~\text{K}$, essentially all collisions are short-lived, but the situation is different inside a microcavity with sufficiently large coupling strength.
 
\par Collision time histograms for simulations of 512 atom pairs, totaling over 40 runs, are presented in Figure \ref{histogram_b}(a-1). This figure confirms that all collisions in free space are ultrafast and never lasted longer than $t_a = 10$ ps in our simulations. We also note that even the longest lived collision in free space contained no signature of oscillatory Ar-Xe motion. Henceforth, for the sake of convenience, we set $t_a$ as a threshold time for association events, i.e., Ar-Xe radiative association is taken to happen any time the collision time is greater than $t_a$ (see Supporting Information for more in-depth discussion). A summary of the statistical properties of simulations involving transient Ar-Xe complex formation under various conditions is provided in the first two rows of Table \ref{table_summary}. 

As expected, the number of association events increases with a higher dipole scaling factor $\gamma$, as shown in Fig. \ref{histogram_b} and Table \ref{table_summary}. This trend indicates that the formation of transient Ar-Xe complexes is strongly influenced by the energy exchange between light and matter. The increase in the dipole scale $\gamma$ also significantly shifts the distribution of Ar-Xe collision times. For instance, in a microcavity with $\gamma = 20\overline{L}$, the collision time distribution skews leftward compared to $\gamma = 10\overline{L}$. Statistically, as shown in the first two rows of Table \ref{table_summary}, the average lifetime of complexes in the microcavity with $\gamma = 20\overline{L}$ is nearly half compared to the case where $\gamma = 10\overline{L}$. This finding highlights a trade-off between the rate of Ar-Xe complex formation and their stability. As the rate of radiative complexes formation rises with $\gamma$, so does the possibility of radiative dissociation \cite{dunbar1994kinetics, dunbar1998activation, dunbar2004bird, suyabatmaz2024polaritonic}.

\begin{figure}
    \centering
    \includegraphics[width=1\linewidth]{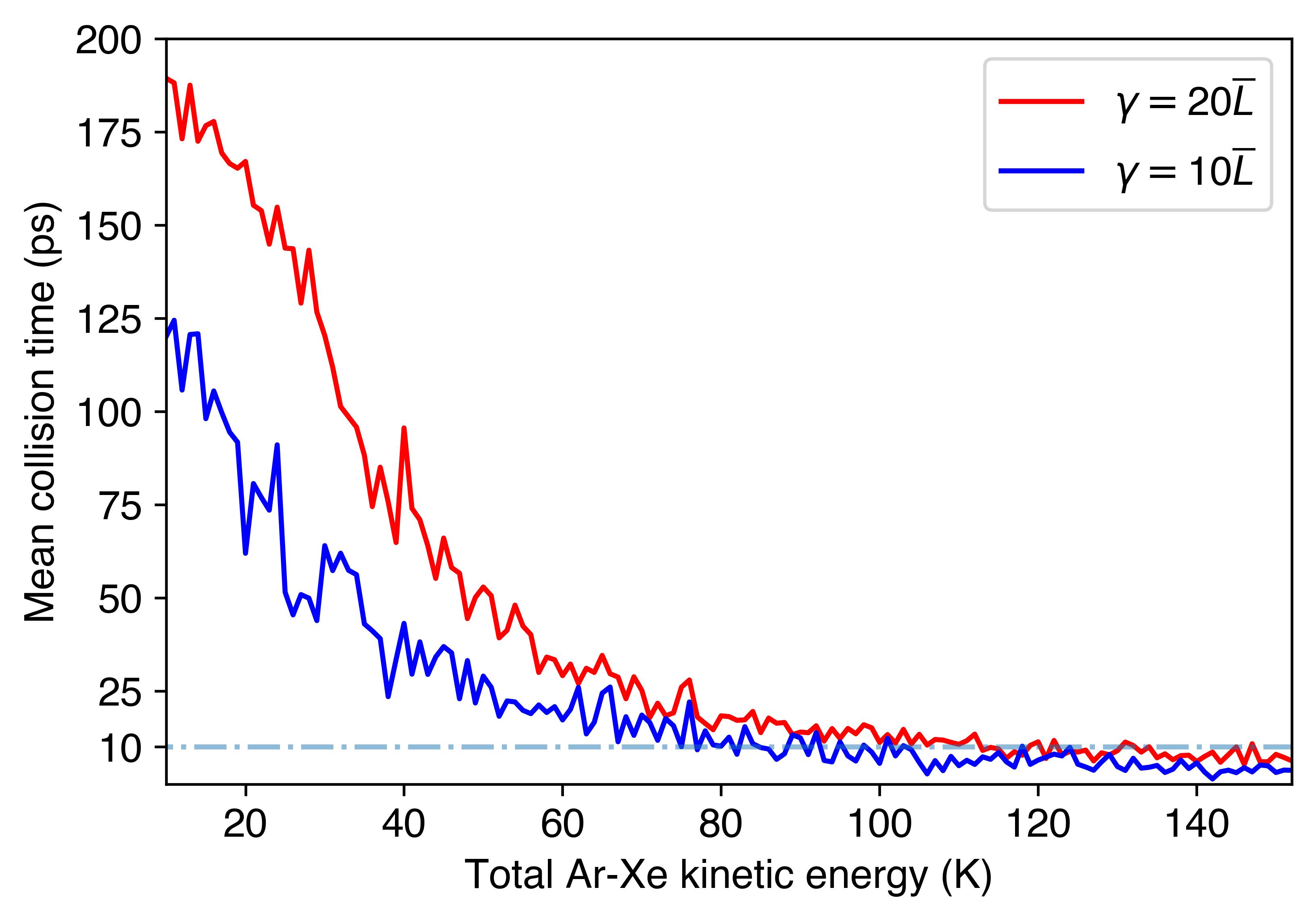}
    \caption{Mean (head-on) collision time of a single Ar-Xe pair coupled to a microcavity radiation field in thermal equilibrium as a function of total kinetic energy. The horizontal dash-dotted line at 10 ps indicates our threshold for radiative association (see Supplementary Information and text). The mean lifetime of Ar-Xe complexes decreases rapidly with increasing collisional energy. At low energies, transient Ar-Xe complexes persist significantly longer in a microcavity with $\gamma =20\overline{L}$ compared to one with $\gamma = 10\overline{L}$, due to the greater probability of radiative losses and subsequent trapping within the Ar-Xe potential well during collisions in the stronger coupling regime.}
    \label{collision_time_b}
\end{figure}

The radiative association may be viewed to arise from the non-conservative character of the energy conversion between the atomic kinetic and potential energy in the presence of sufficiently strong interaction with the EM field. In particular, mechanical energy leaks during the collision are essentially due to the emission of EM radiation. As a result, Ar-Xe complexes may find themselves with insufficient kinetic energy to escape the molecular potential well. The reverse process — where an Ar-Xe complex absorbs field energy and escapes the potential well —  is also observed and eventually leads to dissociation of the transient noble gas complex. Representative trajectories of Ar-Xe association in the presence of the enhanced microcavity radiation field show reversible shuttling of energy between the field and the complex over its lifetime,  placing the dynamics in an intermediate regime, between weak and strong light-matter coupling.

In Figure \ref{collision_time_b}, we provide an estimate of the average lifetime of collinear Ar-Xe collisions as a function of collisional kinetic energy as obtained from the simulation of 100 single head-on collisions between Ar and Xe atoms with equal kinetic energy and with the radiation field at the same temperature. The results indicate a general trend of decreasing lifetimes for Ar-Xe transient complexes as the total kinetic energy increases. This trend is somewhat consistent with previous investigations using quantum scattering theory \cite{cederbaum_making_2024} and radiative association studies \cite{szabo_polyatomic_2023, nyman2015computational}, both of which have shown that the formation of a molecule generally becomes less likely as the collisional energy increases. Similarly, Ar-Xe complexes formed under greater light-matter interaction strength conditions ($\gamma = 20\overline{L}$) exhibit longer lifetimes at low energies due to the increased propensity of emission events facilitating radiative association.

\subsection{Ar-Xe Complex Stabilization in an Ultracold Microcavity}
\begin{figure}
    \centering
    \includegraphics[width=1.0\linewidth]{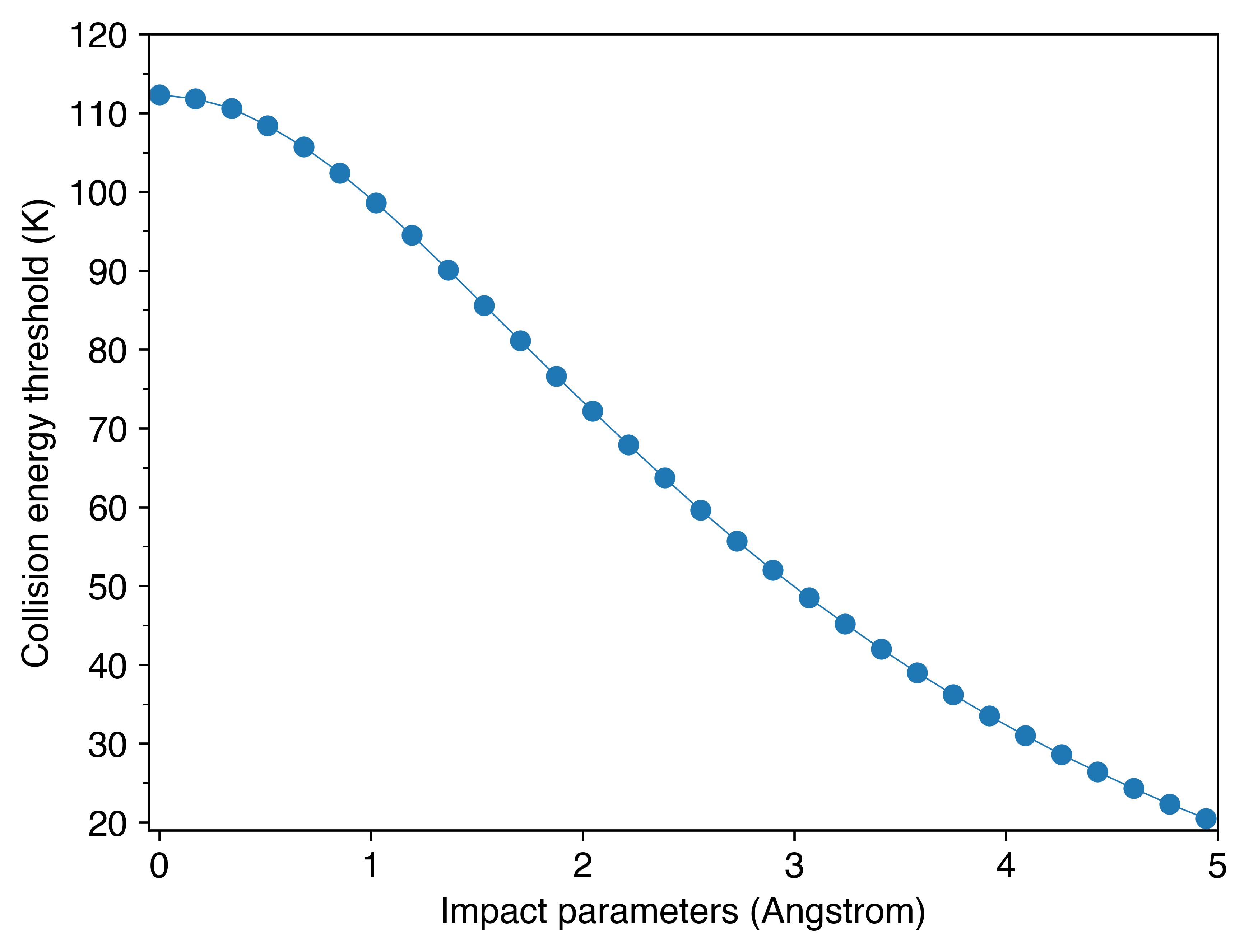}
    \caption{Collision kinetic energy threshold for Ar-Xe radiative association in a 0K microcavity as a function of impact parameter. Initial conditions were set to give Ar and Xe equal kinetic energy. These results indicate that more head-on collisions (smaller impact parameters) increase the likelihood of Ar-Xe complex formation due to shorter interatomic distances at closest approach. The reduced separation produces stronger dipole fluctuations and enhanced interactions with the radiation field, leading to emission by the diatomic system that enables the formation of transient Ar-Xe molecules.}
    \label{energy_threshold_vs_h}
\end{figure}

%Since the dissociation of Ar-Xe transient complexes is driven by the absorption of EM field energy, it is conceivable that lowering the field temperature relative to the molecular would naturally extend the lifetime of the Ar-Xe complex.

Since the dissociation of Ar-Xe transient complexes is driven by the absorption of EM field energy, it is conceivable that the lifetime of the Ar-Xe complex can be extended by lowering the field temperature relative to the atoms. 
This could be implemented by initially reducing the thermal microcavity EM energy density by coupling it to an ultracold bath, and subsequently inserting the warmer gas-phase mixture into the resonator.  In this section, we examine this hypothesis by considering an extreme scenario where Ar-Xe collisions occur in a microcavity where the initial field amplitudes are set to zero, effectively simulating a 0 K resonator. We focus exclusively on strong coupling conditions with a dipole scaling factor of $\gamma = 20\overline{L}$, where radiative association and dissociation are pronounced. 

Limiting our considerations for simplicity to a single colliding pair in a microcavity, we find Ar-Xe collisions in an ultracold resonator lead to radiative association and indefinitely long lifetimes when the initial kinetic energy is lower than a threshold set by the collision impact parameter (the initial perpendicular offset between the relative trajectory of the colliding atoms) \cite{newton2013scattering}. For example, a collinear collision (with vanishing impact parameter) in an empty microcavity with $\gamma = 20\overline{L}$ can result in stable Ar-Xe complexes if the collision kinetic energy is below approximately 112K. This energy threshold decreases as the impact parameter increases, as illustrated in Fig. \ref{energy_threshold_vs_h}. This finding is consistent with previous work \cite{cederbaum_making_2024, szabo_polyatomic_2023, nyman2015computational} and can be simply explained by the shorter Ar-Xe distance of closest approach in collisions with lower impact parameter (i.e., closer to head-on collisions), which leads to stronger vibrational dipole fluctuations, and greater infrared emission probability during a collision.

\begin{figure}
    \centering
    \includegraphics[width=1.0\linewidth]{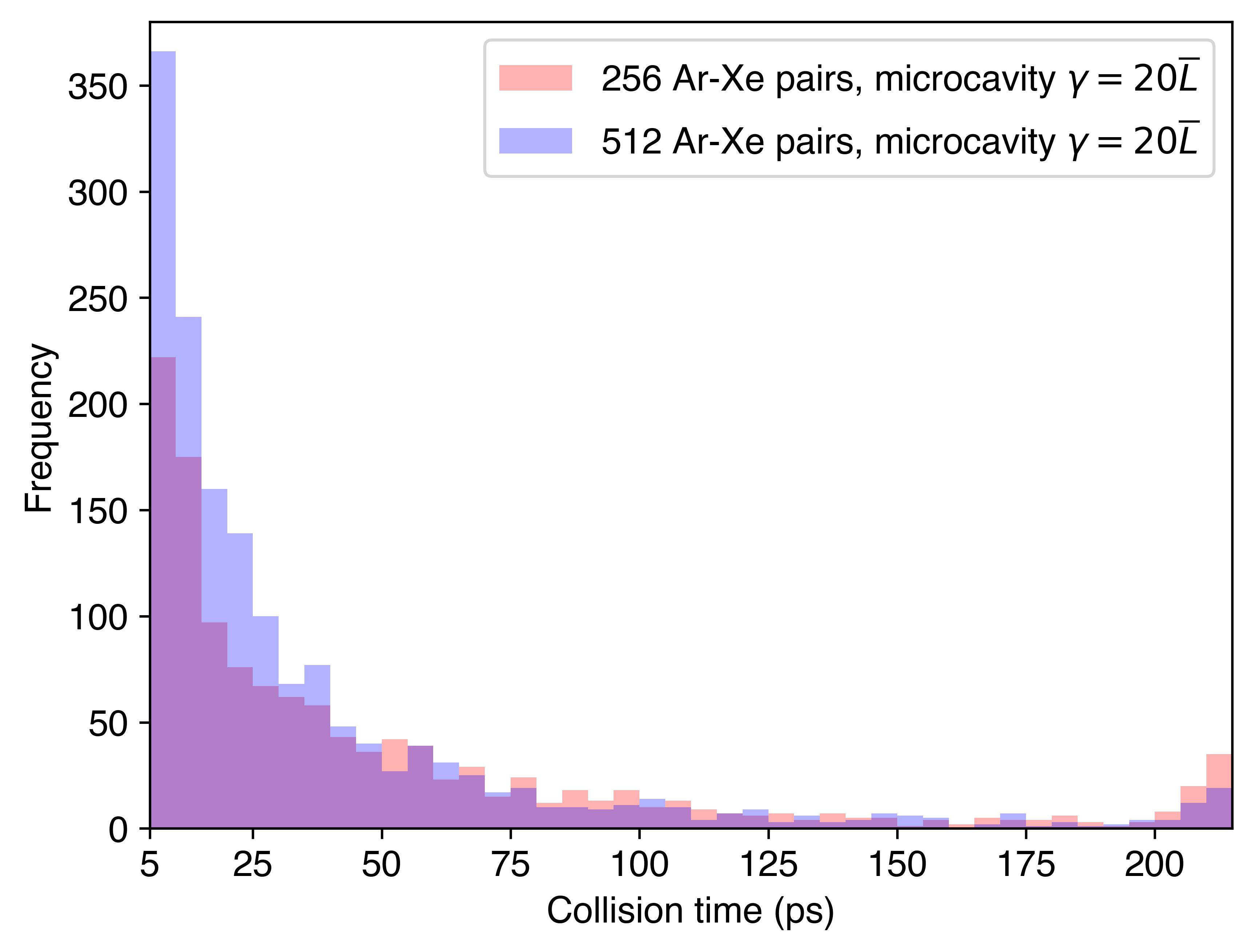}
    \caption{Histograms showing the distribution of collision times exceeding 5 ps for simulations containing 256 and 512 Ar-Xe pairs in a 0K microcavity, with 5 ps bin sizes. The distribution becomes more left-skewed as the number of Ar-Xe collision pairs increases but remains generally less left-skewed than simulations in the 100K cavity. This difference suggests that Ar-Xe complexes formed in the 0K microcavity exhibit greater stability.}
    \label{histogram_z}
\end{figure}

This result shows that stable Ar-Xe complexes can form in a low-temperature microcavity with sufficiently large light-matter coupling strength. However, in Ar-Xe mixtures with multiple pairs, even if the system is sufficiently dilute that dissociation by collisions is suppressed, emission of radiation by the various Ar-Xe pairs will ultimately lead to thermalization of the radiation field and eventually promote complex dissociation. To investigate this, we simulated collisions of 256 and 512 Ar-Xe pairs coupled to a zero-amplitude microcavity with multiple replicas, totaling 20,480 collisions. Collision duration histograms limited to those lasting 5 ps or longer are shown in Figure \ref{histogram_z}, with key statistics summarized in Table \ref{table_summary}. The computed spectra, which closely resemble those in Fig. \ref{cavespec_b}, are included in the Supporting Information.

Fig. \ref{histogram_z} suggests that the number of associations is proportional to the number of Ar-Xe pairs. This effect can be simply attributed to the increase in collision-induced emission processes occurring in the system with a greater number of Ar-Xe pairs, which leads to larger field amplitudes and, hence, effectively stronger light-matter coupling (Eq. \ref{eom_field}).  We also note that the collision time distribution shifts leftward as the coupling strength increases, which is consistent with the findings from the previous section. 

Nevertheless, lowering the microcavity field energy statistically enhances the stability of Ar-Xe complexes. When comparing Fig. \ref{histogram_b}b-2 with Fig. \ref{histogram_z}b, which differ only in field temperature, the latter shows a reduced leftward skew and a more pronounced right tail. This observation is supported by the statistics in the second and last rows of Table \ref{table_summary}, indicating that the 0 K resonator produces a comparable rate of complex formation to the 100 K microcavity but results in complexes with longer mean lifetimes and more complexes surviving beyond 200 ps. This trend is also supported by the more pronounced increase in the final microcavity temperature of the 0 K cavity (Table \ref{table_summary}) compared to 100 K, suggesting that transient complex formation via emission is more prevalent in the former, as expected from their distinct initial conditions.

The overall collision statistics in Table \ref{table_summary} may also be interpreted as follows. Among the considered scenarios, long-lasting Ar-Xe complexes are most likely to form in a microcavity with $\gamma = 20\overline{L}$ and 512 collisions. This could correspond to the case where a microcavity dissipates all its energy after every 512 Ar-Xe collisions. However, if the resonator instead resets its amplitude to zero after only 256 collisions, the number of complexes persisting beyond 200 ps roughly doubles. This observation suggests that higher photon dissipation rates in a lower-temperature environment can enhance molecular formation \cite{cederbaum_making_2024}. Conversely, if the microcavity thermalizes with its environment until the field reaches 100 K, the number of long-lived complexes decreases drastically due to the greater likelihood of thermal radiative dissociation at higher temperatures. 

\subsection{Ar-Xe Vibrational Strong Coupling}\label{subsec_polariton}

\begin{figure}
    \centering
    \includegraphics[width=1.0\linewidth]{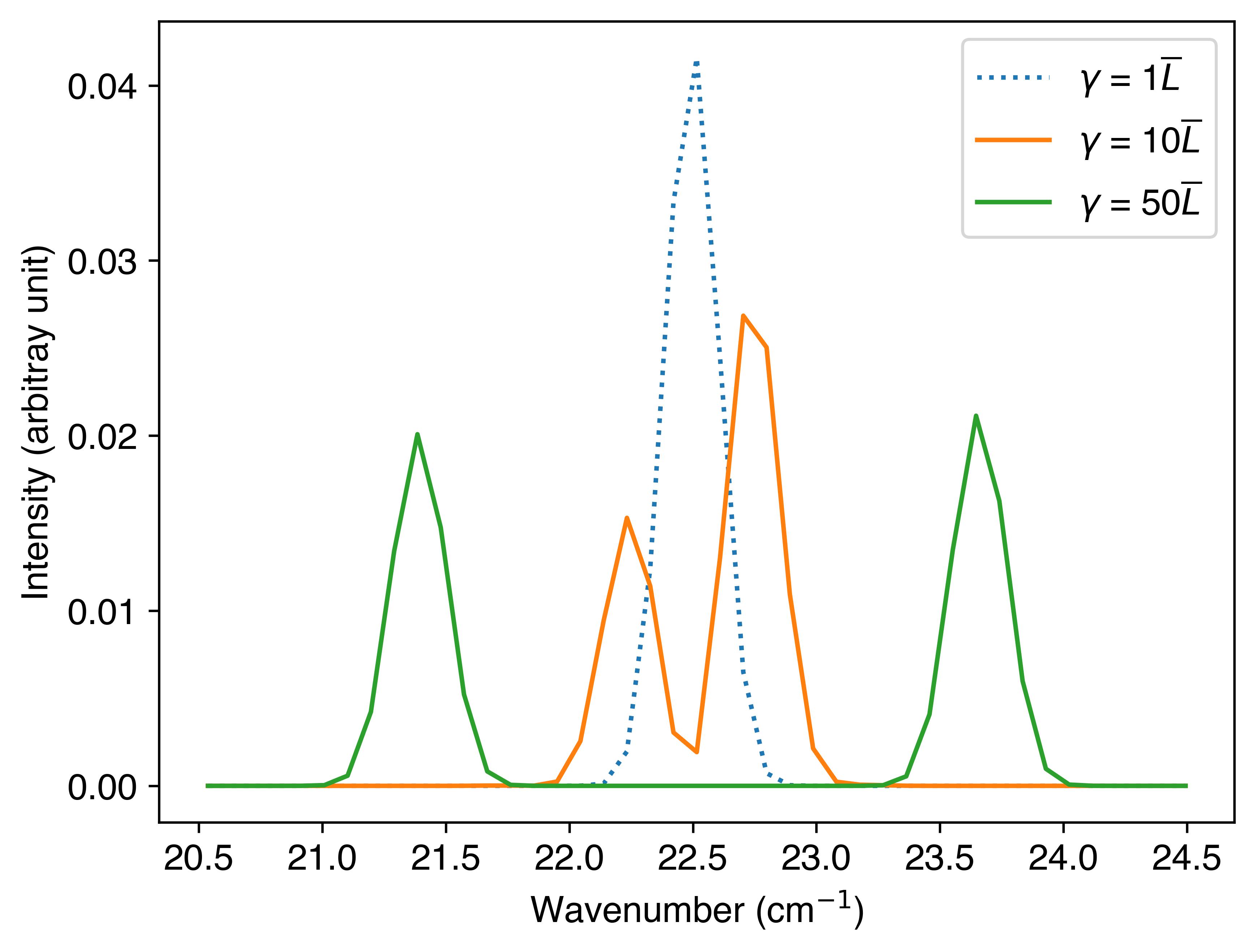}
    \caption{Spectra of a single Ar-Xe complex coupled to a single-mode microcavity field with a frequency equal to the harmonic natural frequency corresponding to the Ar-Xe potential well. The Ar-Xe complex is initialized with zero kinetic energy and distance near the minimum of its potential well, yet a clear Rabi splitting indicating polariton formation is only observed when the dipole function $\mu(r)$ is rescaled by a factor of $\gamma \geq 10 \overline{L}$.}
    \label{polariton}
\end{figure}
Before concluding, we consider the possibility of polariton formation via vibrational strong coupling between Ar-Xe and infrared microcavities. We initialize a simulation with a single Ar-Xe complex with zero initial velocity and distance near the minimum of the potential well. This configuration is coupled to a single microcavity mode at the Ar-Xe harmonic resonance frequency, and the reported spectrum consists of the Fourier transform of the autocorrelation function of the dipole velocity \cite{bornhauser_intensities_2001, praprotnik_molecular_2005}. The results obtained with various dipole scaling factors $\gamma$ are presented in Fig. \ref{polariton}. 

Figure \ref{polariton} shows the emergence of a Rabi splitting and vibrational coupling with Ar-Xe complexes, which requires the enhanced coupling strengths employed above in our analysis of microcavity effects on Ar-Xe complexes. In the classical picture, polariton formation arises from the periodic absorption and emission of electromagnetic energy by the molecular oscillator. This interpretation is supported by previous work \cite{zhu1990vacuum,kavokin2017microcavities,li_cavity_2020}. 
However, we note that observing polariton signatures in Ar-Xe complexes in a real mixture under strong coupling is highly improbable, as any formed Ar-Xe complexes are metastable and have a very short lifetime. These features lead to broad vibrational optical lineshapes that prevent the conditions for strong coupling from being achievable except potentially at ultralow temperatures approaching 0 K. 

\section{Summary and Conclusions}\label{sec_conclusion}

We implemented a classical molecular electrodynamics model in the Coulomb gauge to investigate collisional-induced spectra and the association dynamics of a mixture of Ar and Xe atoms in a multimode EM field. Our free space simulations provide collision-induced emission spectra that match well with experiments. The broad lineshape of the collision-induced emission prevents the regime of strong light-matter interactions from being achieved in a microcavity except under extremely low temperatures and special initial atomic configurations. Nevertheless, in a microcavity with sufficiently large light-matter coupling, we find that collision-induced infrared spectra are mainly changed due to radiative association and dissociation processes, which become favored relative to free space and induce a substantial change in the tails of the transient complex lifetime distribution. 

Our detailed analysis of the collisional dynamics and stability of the transient Ar-Xe complexes in a microcavity with enhanced radiative coupling relative to free space suggests that tuning the confined EM field temperature (e.g., via embedding the empty microcavity in an ultracold environment) can significantly modulate the lifetime distribution and enhance the stability of weakly bound Ar-Xe complexes. 

While our quantitative studies have focused on how enhanced light-matter interactions affect the association of weakly interacting atoms in confined environments, we expect our main conclusions to have broader implications. Specifically, multimode photonic resonators with enhanced light-matter coupling may be employed to manipulate the stability of weakly bound complexes across various systems, regardless of whether polariton formation occurs, as long as sufficiently large light-matter interaction strengths can be achieved.

Finally, our theoretical analysis offers a potential mechanism for microcavity-induced change in chemical reaction kinetics. Consider the following simple reaction with an intermediate complex:
\begin{align*}
    A-B + C \rightleftharpoons [A-B \dots C] \rightleftharpoons A + B-C.
\end{align*}
Field-induced change in the $B...C$ association rate and $A-B$ dissociation would indeed modulates the reaction kinetics. This suggests future multimode investigations of the radiative association mechanism on multiple-step polyatomic reactions \cite{moiseyev2024complex}.

\section*{SUPPLEMENTARY MATERIAL}

\par The supplementary material contains derivations of the implemented Coulomb gauge Lagrangian and collision-induced lineshape, tables of parameter values for the Lennard-Jones potential and dipole function, and additional simulation details. Additional figures that support our findings are also included.

\section*{DATA AVAILABILITY}

The data that supports the findings of this study are available within the article and its supplementary material. The source code for this project is available at \href{https://github.com/RibeiroGroup/Molecular-Dynamics-for-CIS}{https://github.com/RibeiroGroup/Molecular-Dynamics-for-CIS}.

\acknowledgments{R.F.R acknowledges support from NSF CAREER award Grant No. CHE-2340746, and startup funds provided by Emory University.}

\bibliography{aipsamp}% 

\end{document}